# Prediction-Based Data Transmission for Energy Conservation in Wireless Body Sensors

Feng Xia, Zhenzhen Xu, Lin Yao, Weifeng Sun, Mingchu Li
School of Software
Dalian University of Technology
Dalian 116620, China
f.xia@ieee.org

*Abstract* — **Wireless body sensors are becoming popular in healthcare applications. Since they are either worn or implanted into human body, these sensors must be very small in size and light in weight. The energy consequently becomes an extremely scarce resource, and energy conservation turns into a first class design issue for body sensor networks (BSNs). This paper deals with this issue by taking into account the unique characteristics of BSNs in contrast to conventional wireless sensor networks (WSNs) for e.g. environment monitoring. A prediction-based data transmission approach suitable for BSNs is presented, which combines a dual prediction framework and a low-complexity prediction algorithm that takes advantage of PID (proportional-integral-derivative) control. Both the framework and the algorithm are generic, making the proposed approach widely applicable. The effectiveness of the approach is verified through simulations using real-world health monitoring datasets.**

## I. INTRODUCTION

Over years the field of wireless sensor networks (WSNs) has achieved huge advancements in various aspects including fundamental theory, key technologies, and real-world applications. With an increasing number of wireless sensors becoming commercially available, WSNs have been enabling a broad range of novel applications, among which human health monitoring is a typical example. Continuous health monitoring is a key technology for realizing the transition of current health care systems to more proactive and affordable healthcare, especially for the elderly. This service has tremendous potential to help address the world-scale challenge of population aging. A WSN used in the context of health care is usually named under body sensor network (BSN) [1,2], a relatively new term with a history of only a few years.

From a historical perspective, however, body sensors are an established technology that has emerged more than one hundred years ago [3]. One notable milestone is the Allbutt's invention of the clinical thermometer used for taking temperature of a person in 1867. During the past century, body sensors of various types and with disparate functionalities have been put into use, becoming smaller and smaller in size. Thanks to recent technological developments in electronics and wireless communications, especially the personal area networking standards such as IEEE 802.15.1 (Bluetooth), 802.15.4 (Zigbee), and 802.15.6 (the work-in-progress body area network protocol), many physiological sensors have transformed into wireless sensors in possession of sensing, computing, and (wireless) communication capabilities. These on-body or in-body wireless sensors, capable of communicating with other devices, make it possible to realize continuous monitoring of patients in hospital and long-term health care for the aged and/or the disabled in their homes.

A BSN consists of a number of wireless body sensor nodes, which measure diverse physiological phenomena of a human body, such as blood oxygen, electrocardiography (ECG), electroencephalography (EEG), electromyogram (EMG), central venous pressure (CVP), respiratory impedance, pulmonary arterial pressure (PAP), and temperature. Generally, there is an additional base station (also called *sink*) possibly functioned by a mobile phone or a PDA (personal digital assistant), etc. Healthcare systems built upon BSNs can potentially transform how people's health is monitored, how chronic illnesses are treated, and how the damages of acute events are minimized. For instance, health data collection is traditionally conducted intermittently, e.g., once every several days via a doctor visit. As a consequence, the time points of health monitoring are very limited. On one hand, abnormal, even life-threatening events might be unobserved. On the other, the value of the data collected as for aid of diagnosis is restricted due to undersampling [2]. In contrast, BSNs allow to closely monitor, in a continuous manner, the physiological states of an individual, and to provide real-time warnings of abnormalities, both helping address the drawbacks of the traditional health data collection approach.

To realize the great potential of BSNs in practice, however, a number of obstacles associated with size, lifetime, compatibility, privacy and security must be tackled. In particular, BSNs have to operate on extremely limited energy budget. In order for the users to feel comfortable and acceptable to wear or even implant body sensors, they should be made as small as possible. This imposes critical constraints on the size of the batteries. Since a battery's capacity is in most cases proportional to its size, the battery energy available to wireless body sensors becomes a very scarce resource [1-4]. In addition, it would be inconvenient and unpleasant, if not impossible, for the users to recharge or change the battery of a

This work is partially supported by Natural Science Foundation of China under Grant No. 60903153.

body sensor, especially when it is implanted. While current sensor nodes (also called *motes*) often feature short lifetimes (on the order of hours or days) and relatively bulky batteries (e.g. AA battery), wireless body sensors are expected to operate over months or even years. Therefore, energy conservation at runtime becomes a first class design issue for BSNs.

In the realm of WSN, it has been well recognized that data transmission is generally much more expensive in terms of energy consumption than data processing and sensing. The energy spent in transmitting one bit is hundreds to thousands times that spent in executing one instruction [5]. Given this fact, it is possible to achieve overall energy savings by making tradeoffs between computation and communications. Nevertheless, this should be done with adequate carefulness because wireless sensors are always limited in computational power, which is more pronounced for body sensors.

The objective of this paper is to save the energy of body sensors as much as possible while preserving the quality of the health monitoring application. Energy conservation is achieved by reducing the amount of data transmitted over the wireless networks, i.e., only a subset of the sampled data produced by the body sensors will be delivered to the base station. Intuitively, this would sacrifice, to some extent, the accuracy of the information captured at the base station. To overcome this potential shortcoming, the subset of delivered data is selected in a way that the original samples could be reconstructed at the base station with a guaranteed quality. To this end, the following two issues need to be addressed: (1) for every sample at the sensor node, how to decide whether or not to send it to the base station, and (2) at the base station, how to reconstruct a sample not sent by the sensor node while satisfying user-defined accuracy requirements? In this paper, we take advantage of the *dual prediction* [5,6] approach to address these problems in the context of BSNs. The major contributions of this paper are as follows.

- We introduce the dual prediction framework as an effective solution for energy conservation in wireless body sensors. Although the dual prediction approach has been studied and applied to WSNs previously (e.g [7-9]), to the best of our knowledge, we are the first to apply this approach to the area of BSNs, which are quite different in features from conventional WSNs for e.g. environmental monitoring (see Section II). Due to the unique characteristics of body sensors and the specific requirements of healthcare applications, the effectiveness of existing dual prediction methods need to be re-assessed when applied to BSNs.

- We propose an efficient and generic prediction algorithm that borrows the concept of proportional-integral-derivative (PID) control [10,11], the most popular design technique in control engineering. The original PID control algorithm is adapted to be used for prediction of sensor measurements. The proposed algorithm is lightweight with low complexity, which makes it well suited for running on body sensors with limited computational capabilities. Thanks to the well-established PID control theory, our prediction algorithm is model-free and relatively insensitive to changes in its coefficients. Therefore it can be utilized in a broad range of applications, e.g., for monitoring various physiological parameters simultaneously.

- We conduct simulations using traces from real-world datasets to evaluate the performance of the proposed approach. The results show that in comparison with the traditional data transmission method, our approach can reduce the energy consumption significantly (by 59% to 99% in the experimentations), and at the same time, guarantee required information quality in terms of pre-configured data error bound.

The rest of the paper is organized as follows. Section II gives some background for this work. The unique characteristics of wireless body sensors are described in contrast to conventional wireless sensors for environmental monitoring. Some existing works closely related to this paper are reviewed. In Section III, we formulate the problem to be solved. Sections IV and V present the framework and the algorithm for the proposed prediction-based data transmission approach, respectively. We examine the performance of the proposed approach via trace-based simulations in Section VI. Preliminary yet promising results are provided. Finally, Section VII concludes the paper.

## II. BACKGROUND

### A. Body Sensor Networks

BSNs constitute a special category of WSNs. Like conventional wireless sensors, body sensors are typically comprised of four main components, i.e., a sensing unit, a processing unit, a communication unit (i.e. transceiver), and a power supply unit. However, due to the various specific requirements imposed by healthcare applications, BSNs are featured by many unique design questions demanding new research efforts [2]. In this subsection, we will examine the characteristics of BSNs in comparison against conventional WSNs. For simplicity, environmental monitoring is taken as an example of target applications of the conventional WSNs in this context.

A BSN often contains a certain number of body sensor nodes and a base station, as shown in Fig. 1. Each sensor node, either wearable or implanted into the human body, monitors a certain physiological parameter. It transmits (via a wireless link) sampled data to the base station, which is responsible for collecting the data from all sensor nodes and in some cases forwarding them to a remote data center on the Internet through a WLAN (wireless local area network) or cellular network.

A comparative study of BSNs and conventional WSNs with respect to some critical features is given in Table I. Unsurprisingly, BSNs and environment monitoring WSNs share many characteristics and application requirements. For example, both feature resource constraints and demand energy-efficient operations, security protection, QoS (quality of service) provisioning, and so on. Interesting to note is that there exist a considerable number of key differences between these two systems [1,12]. For instance, while there are typically large numbers of sensor nodes in environmental monitoring, a typical BSN has only a few body sensors. This is because the physical area for deployment of physiological sensors is restricted to the human body, and furthermore, they must be properly placed at certain particular points to achieve effective measurements. Few sensor nodes make it hard to exploit redundancy in BSNs, which brings considerable challenges to body sensor networking.

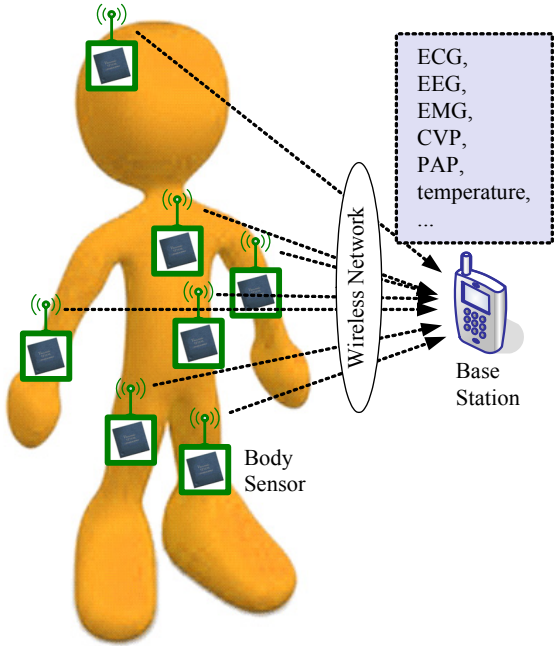

Figure 1. Architecture of a typical body sensor network.

TABLE I. COMPARISON OF BSNs AND CONVENTIONAL WSNs

| Issue | BSN | Conventional WSN |
|---|---|---|
| Scale | A few (less than 10) nodes around human body, without redunduncy | Numerous (10s to 1000s) nodes over a large area, with high degree of redundancy |
| Topology | Star | Hybrid/diverse |
| Node variety | Heterogeneous | Most are homogeneous |
| Physical features of nodes | Very small, lightweight, wearable or implantable, ergonomic, unobtrusive | Not significant constraints |
| Resources | Extremely limited | Limited |
| Energy efficiency | Extremely energy-efficient | Energy-efficient |
| Privacy | Protecting sensitive personal data is highly critical | Not a significant constraint |
| Communication | Single-hop, QoS-enabled | Multi-hop, less QoS-sensitive |
| Example of nodes | Sensium | MicaZ, IMote2 |

Yet another noteworthy difference is that unlike conventional WSNs where most (if not all) sensors are homogeneous, body sensors are normally heterogeneous. This is quite intuitive in consideration of the requirements on body sensor types and wearability, which could vary significantly from one application (corresponding to the monitoring of a particular physiological parameter) to another. In addition, a star topology is a natural choice for BSNs since all nodes are located near the human, though mesh topology could be integrated in some situations. The data transmissions in BSNs are generally single-hop, which is in contrast to environmental monitoring with WSNs employing multi-hop communications.

As mentioned previously, both of BSNs and conventional WSNs suffer from resource constraints. However, this problem is much more pronounced in the case of BSNs. Minimization of size and weight is of paramount importance for body sensors because they must be as noninvasive as possible to achieve social acceptance [1,12]. Being smaller in size and lighter in weight imposes further constraints on the available resources such as clock frequency, memory, and battery energy. For example, conventional wireless sensor nodes often use a pair of AA batteries that provide a few Watt-hours of energy. The Sensium chip from Toumaz, a wearable body sensor node released recently, operates on paper-thin printed battery characterized by a capacity of about only 20 mWatt-hours [4].

*B. Related Work*

Energy conservation for WSNs has been extensively studied in recent years. A recent survey on this field is given in [5]. Among many other solutions, data reduction is an effective technique to achieve energy-efficient data acquisition. The basis idea of data reduction schemes is to reduce the amount of data transmitted over the network such that the energy spent for communications is cut down. Since communications are commonly the most expensive operation in terms of energy consumption, data reduction could significantly prolong the lifetime of wireless sensors. In the literature, *data prediction* is one of the data reduction schemes that have been exploited most widely.

Goel and Imielinski [13] proposed a prediction-based monitoring paradigm called PREMON based on the following two observations. The first is that spatial correlation of sampled data can be exploited to estimate the sample at a sensor based on the knowledge of samples of sensors around it and their past values. Secondly, a snapshot of a sensor network can be regarded as an image and consequently the evolution of samples may be visualized as a *sensor movie*. Exploring the concept of MPEG, PREMON generates a prediction model at the base station and sends it to the sensors, in a periodic manner. A sensor transmits its sample only when it derivates from the predicted value of the model by more than a user-defined threshold.

Jain *et al.* [7] presented a dual Kalman filter scheme for minimization of communication overhead. In the work, the authors exemplified the concept of *dual prediction*, and advocated the use of the Kalman filter for stream filtering. Chu *et al.* [9] employed the same idea as [7] to minimize communication from sensor nodes to the base station. In addition to dual prediction, the work also exploited spatial correlations of sampled data from different sensors. The presented solution uses replicated dynamic probabilistic models, with a framework quite similar to the dual Kalman filter scheme. The pair of probabilistic models, one at the base station and another distributed in the network, is updated whenever they are considered invalid.

As the sampled values of a physical variable in general constitute a time series, it is easy to understand that time series forecasting consists in a typical method used for data prediction in WSNs. For instance, the PAQ solution proposed by Tulone and Madden [8] uses autoregressive models generated at each sensor nodes to predict local samples. Sensors send their local models to the base station where the sampled values of each sensor are predicted using the corresponding model. Whenever necessary, a sensor will send information about outliers and updated parameters of the model to the base station. In most cases, the model used for data prediction is fixed *a priori*, though the model parameters may be adapted at run time. As such, a bad choice of prediction model e.g. due to lack of adequate and accurate *a priori* knowledge could lead to poor prediction performance. To address this issue, Borgne *et al.* [6] proposed an adaptive model selection strategy that allows sensors to autonomously choose at run time the best model (in terms of statistical performance) out of a set of candidate models.

In [14], Santini and Römer presented a data prediction approach that does not depend on *a priori* knowledge to correctly model the sampled data. In the work, the well-established least-mean-square (LMS) algorithm is adopted. Lazaridis and Mehrotra [15] explored the idea of combining data compression with prediction. Several issues related to data prediction, including model selection, were discussed therein.

Although a considerable number of methods have been developed for data prediction, all of the above mentioned works deal with conventional WSNs for e.g. environmental monitoring. The applicability of these methods to BSNs needs to be re-assessed in consideration of the differences between BSNs and conventional WSNs, as discussed in the previous sub-section. For example, a Kalman filter is often too computationally expensive to run on a body sensor due to its extremely limited data processing capability. For typical BSNs, it is impossible to exploit spatial correlation among sampled data because there is very little or no sensor redundancy. This makes many existing methods unsuitable for BSNs. In addition, the diversity of physiological signals necessitates that the employed prediction model must be applicable to a variety of applications. As far as we know, there exists no previous work that has addressed data prediction based energy conservation of wireless body sensors.

### III. PROBLEM FORMULATION

In this paper we consider a body sensor network with a star topology for health monitoring as shown in Fig. 1. Each body sensor communicates with the base station directly. Without loss of generality, we assume that the sensors are independent of each other, i.e., there is no sensor redundancy. Since we treat each pair of sensor and base station in the same manner, henceforth we will illustrate our approach by focusing on only one sensor (for the sake of simple notation and description). In addition, the following assumptions are made:

- Assumption 1: The sensor and the base station are synchronized in time clock.
- Assumption 2: The computation and communication latency is negligible.
- Assumption 3: There is no packet loss over the network.

Let $x_k \in \mathbb{R}$ be the sampled value of the monitored physiological parameter at time step $k$. Note that we are not concerned with the actual sampling period of the sensor in this work. The time step $k \in \mathbb{N}^+$ does not represent a particular point of time but the sequence of samples. Let $y_k \in \mathbb{R}$ be the recorded vlaue of the physiological parameter at time step $k$, which is stored at both the sensor and the base station in this work. It is held that:

$$y_k = \begin{cases} x_k & \text{if the sample is sent to the base station} \\ \hat{x}_k & \text{otherwise} \end{cases}$$

with $\hat{x}_k$ being the prediction of $x_k$. As in a general form, it is given by

$$\hat{x}_k = f(y_{k-1}, y_{k-2}, ..., y_{k-M}) \tag{1}$$

where $f(\cdot)$ is a mathematical function (also called *prediction algorithm*) and $M \in \mathbb{N}^+$ is the number of previous records.

The prediction error is defined as:

$$e_k = x_k - \hat{x}_k. \tag{2}$$

The objectives of this work are as follows:

- To minimize the energy spent for communications, and
- To satisfy pre-specified quality requirements on data acquisition.

In general, the communication energy dissipation is proportional to the amount of data transmitted over the network. Instead of computing the actual energy consumption, we use the ratio of the number of samples transmitted by the sensor (to the base station) to the number of all samples, which is denoted $r$, as an indirect indicator of energy dissipation. For example, if $N$ out of $k$ samples have been transmitted over the network (up to time step $k$), then $r = N/k$. Traditionally, every sample will be sent to the base station, which yields that $r=1$. For a certain data transmission scheme, the *energy ratio* (also called *normalized energy consumption*, denoted $E$) can be defined as the ratio of the amount of (communication) energy consumed by the scheme to that consumed by the traditional scheme. Therefore, $E \approx r$.

In order to measure the quality of the health monitoring application, we define *data error* (denoted $\tilde{e}$) as the difference between the actual (sampled) value of the physiological parameter and the corresponding recorded value at the base station, i.e.,

$$\tilde{e}_k = x_k - y_k. \qquad (3)$$

Let $\varepsilon \in \mathbb{R}^+$ be the maximum allowable (absolute) data error. It is a user-specified parameter that highly depends on the target application. In this work we assume it is known for each application. Then the application quality requirement can be stated as follows:

$$|\tilde{e}_k| \leq \varepsilon. \qquad (4)$$

As long as (4) holds, any data $y_k$ recorded at the base station will not deviate from the (corresponding) actual value $x_k$ by more than $\varepsilon$.

## IV. FRAMEWORK OF PREDICTION-BASED DATA TRANSMISSION

As mentioned previously, we employ the concept of dual prediction to address the above described problem. It is worthy to note that many dual prediction strategies have been explored in previous works, especially for conventional WSNs. In the following, we will introduce a data transmission framework based on dual prediction that is applicable to wireless body sensors.

According to the traditional data transmission approach, the sensor delivers every sample to the base station once it is available. In contrast, the basic principle of prediction-based data transmission is to selectively transmit sampled data, i.e., only a subset of all samples will be delivered to the base station. For this purpose, the sensor needs to decide, for each sample, whether to send it to the base station. On the other hand, when the sample is not received, the base station should be able to reproduce an estimate of the sample with the required quality. Both of these issues will be tackled by exploiting dual prediction.

In the framework, two *identical* predictors are introduced, one running at the sensor and another at the base station. The predictor executes the prediction algorithm given by (1) and hence can generate an estimate of the sampled value. At run time, the sensor samples data regularly, and at the same time, makes the prediction. If the *absolute prediction error* is bigger than the error budget $\varepsilon$, the sample will be delivered to the base station and simultaneously stored at the sensor. Otherwise, the sensor will discard the sampled value and store the prediction instead, i.e., $y_k = \hat{x}_k$. Accordingly, the condition for data transmission can be expressed as:

$$|e_k| > \varepsilon. \qquad (5)$$

The workflow of the sensor is described in Fig. 2. It can be seen that the previous $M$ samples need to be stored temporarily in order for the sensor to make the prediction. This inevitably introduces some memory burden. Because body sensors have very limited memory (typically less than 20 KB) [1], the value of $M$ should be as small as possible. In addition, the complexity of the prediction algorithm should be as low as possible to make it computationally feasible for a body sensor.

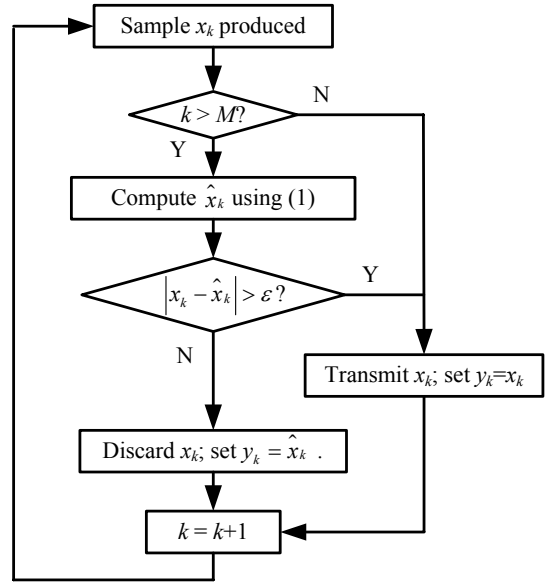

Figure 2. Workflow of the sensor.

The base station has to maintain the required quality (or accuracy) of the data to be recorded even when the sensor decides not to send the sample. In this situation, for example, if the sensor does not deliver $x_k$, the base station will activate the predictor to produce the estimate $\hat{x}_k$ and then records it. The workflow of the base station is illustrated in Fig. 3.

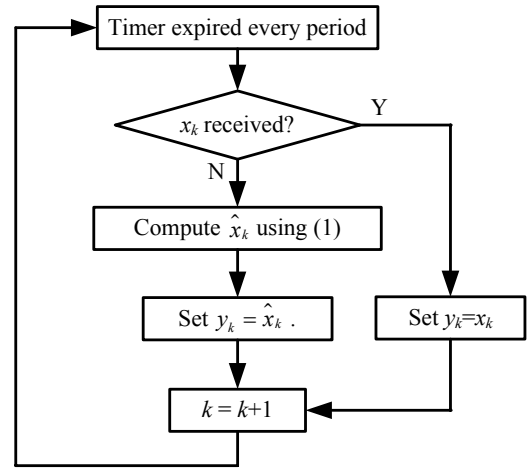

Figure 3. Workflow of the base station.

It is quite easy to examine whether the above described strategy can satisfy the application quality requirement given by (4). Since a certain sample, say $x_k$, will be either delivered or discarded, we will discuss these two situations one by one. Consider first the case in which $x_k$ is delivered. This implies that $|e_k| > \varepsilon$. Because the base station will receive the actual value, i.e., $y_k = x_k$, the data error $\tilde{e}_k = x_k - y_k = 0$. It is apparent that the condition (4) is met. If $x_k$ is discarded by the sensor, we have $|e_k| \leq \varepsilon$ in accordance with the data transmission principle given by (5). In this case, the recorded value at the base station will be $y_k = \hat{x}_k$. Consequently, $|\tilde{e}_k| = |x_k - y_k| = |x_k - \hat{x}_k| = |e_k| \leq \varepsilon$, meaning that (4) is held. In summary, the above data transmission strategy is able to guarantee the required application quality. Some additional properties of our framework are outlined below:

- The framework does not rely on sensor correlation or redundancy, which is generally unavailable in BSNs. It can be applied to each pair of sensor and base station independently.

- The framework is platform-independent because the data transmission strategy is to be implemented at the application layer, making it suitable for heterogeneous body sensors.

- The framework does not adopt a prediction model that needs to be updated regularly when it becomes invalid. Transmitting model parameters between the sensor and the base station, in either direction, often yields considerable increase in communication cost. This may cause significant extra energy consumption in body sensors. We avoid this problem by employing an efficient and generic prediction algorithm, which will be presented in the next section.

## V. PREDICTION ALGORITHM

The prediction algorithm $f(\cdot)$ used in the identical predictors is the most important component of prediction-based data transmission. It has been pointed out that the performance of dual prediction schemes largely depends on the prediction algorithm (or prediction model) [6]. In the context of BSN (as opposed to conventional WSN), the unique characteristics of wireless body sensors (see Section II) give rise to some particular requirements on the prediction algorithm. For instance, the prediction algorithm has to be very low in complexity due to the extremely limited resources of body sensors. A BSN application usually needs to monitor, at the same time, multiple physiological parameters of disparate characteristics, such as ECG, blood pressure, and body temperature. This heterogeneity imposes further necessity that the prediction algorithm should be widely applicable across applications, i.e., it must be able to estimate a variety of signals of different types, with adequate accuracy of course.

Inspired by the work by Varma *et al.* [16] who applied the PID algorithm to CPU workload prediction for dynamic voltage scaling, we proposed a PID-based prediction algorithm for energy-efficient data transmission. In the following, we first introduce the basic principle of PID control, and then present the prediction algorithm used in this work.

### A. What is PID?

The term PID (proportional-integral-derivative) refers to a popular control algorithm that was invented one hundred years ago. It is well known in the automatic control community that PID control is very powerful and is capable of solving a wide rage of control problems. Thus far, more than 95% of all real-world industrial control problems are addressed by utilizing PID control [10]. It has been implemented in various systems of different scales and degrees of complexity, from small and simple devices to huge factories equipped with thousands of controllers. This popularity of PID control mainly benefits from its simplicity, clear functionality, wide applicability, and ease of use [11].

A general architecture of PID control is depicted in Fig. 4. The basic role of the controller is to adjust the *manipulated variable u* with respect to the *control error e* such that the *system output* (i.e. *controlled variable*) *y* will approach the *setpoint $r_s$*. The control error is the difference between the setpoint and the actual system output, i.e., $e = r_s - y$.

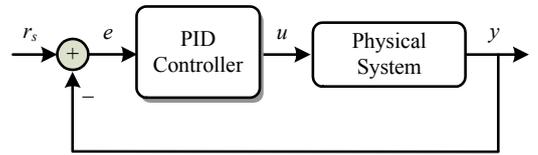

Figure 4. PID control architecture.

As the name implies, a PID controller is comprised of three components: proportional, integral and derivative control. The input/output relation for an ideal PID controller can be written mathematically as:

$$u = K_P e + K_I \int_0^t e(\tau) d\tau + K_D \frac{de}{dt} \qquad (6)$$

where $K_P$ is the proportional gain, $K_I$ the integral gain, $K_D$ the derivative gain and, $t$ the point of time, with the first three called *controller parameters*. From a functionality perspective, the proportional component produces a control action proportional to the control error; the integral component mitigates steady-state errors through low-frequency compensation by an integrator; the derivative component improves transient response through providing the controller with an anticipative ability.

### B. PID-Based Prediction

As described above, the original PID algorithm is devised for feedback control. In order to apply it to data prediction, some modifications must be made. The prediction algorithm used in this work is given by [16,17]:

$$\begin{aligned}\hat{x}_k &= f(y_{k-1}, y_{k-2}, ..., y_{k-M}) \\ &= K_P y_{k-1} + K_I \sum_{i=k-M}^{k-1} \frac{y_i}{M} + K_D (y_{k-1} - y_{k-2})\end{aligned} \qquad (7)$$

Like the PID controller, the above prediction algorithm consists of three components:

- The 'proportional' term $K_P y_{k-1}$ attempts to track the changes in the sampled values quickly by assuming that the current value will remain close, to some extent, to the last record. The dependency between the predicted value and the last record is determined by the proportional gain $K_P$.

- The 'integral' term $K_I \sum_{i=k-M}^{k-1} \frac{y_i}{M}$ considers a longer history of the records than the proportional term. It is based on the point of view that the current value will be close to the mean of the previous $M$ consecutive samples. This term contributes to mitigate the impact of measurement noise but makes the predictor to response more slowly to changes.

- The 'derivative' term $K_D(y_{k-1} - y_{k-2})$ assumes that the rate of change (in the values of the physiological parameter) is likely to remain the same as in the last interval. This term represents a prediction of the change in value.

There are four coefficients in the above prediction algorithm: $K_P$, $K_I$, $K_D$, and $M$. As mentioned previously, the value of $M$ should be as small as possible as long as the resulting performance is good enough. The determination of the other three coefficients seems empirical and difficult. Thanks to the well-established theory of PID control, however, this is not as hard as one might think. In the control community, there exist large groups of tuning method for PID controllers, including analytical methods, heuristic methods, frequency response methods, optimization methods, and adaptive tuning methods. Reviewing these methods is out of the scope of this paper. Interested readers are referred to [10,11] for details.

Several special cases deserve extra attention:

- $K_P = 1$, $K_I = 0$, and $K_D = 0$. In this case, we have $\hat{x}_k = y_{k-1}$, meaning that the predicted current value is equal to the last record. For notational convenience, we name the corresponding algorithm PAST. This algorithm has been explored in many existing works.

- $K_P = 0$, $K_I = 1$, and $K_D = 0$, which lead to $\hat{x}_k = \sum_{i=k-M}^{k-1} \frac{y_i}{M}$. This case corresponds to a typical *moving average* method for time series forecasting. We name it the AVERAGE algorithm.

- $K_P = 1$, $K_I = 0$, and $K_D = 1$. The prediction algorithm given by (7) then becomes $\hat{x}_k = 2y_{k-1} - y_{k-2}$. Since this algorithm yields a linear predictor [18], we name it LINEAR hereafter.

Recall that all these algorithms are special cases of our prediction algorithm. In other words, our algorithm gives a general form of many algorithms available in the literature. In addition, our algorithm is application-independent because it requires no *a priori* knowledge about the sampled data and therefore can be applied to a variety of applications. The simplicity of the algorithm also makes it computationally cheap.

## VI. EVALUATION

### A. Simulation Setup

We conduct simulations based on Matlab to assess the performance of the proposed data transmission approach. To reflect the real application effects of wireless body sensors, five real-world datasets from the Massachusetts General Hospital/Marquette Foundation (MGH/MF) Waveform Database are used in the simulations. These datasets are listed in Table II, with each containing 5000 consecutive samples (selected from the raw data). The MGH/MF Waveform Database, publicly available at PhysioNet [19], is a comprehensive collection of electronic recordings of hemodynamic and electrocardiographic waveforms of 250 patients in critical care units. It represents a broad spectrum of physiologic and pathophysiologic states. Fig. 5 plots the curves of the data used.

TABLE II. DATASETS USED IN THE SIMULATIONS

| Dataset | Source record | $\varepsilon$ |
|---|---|---|
| ECG | mghdb/mgh012 | 0.1 |
| PAP | mghdb/mgh183 | 3.5 |
| ART | mghdb/mgh003 | 5 |
| CVP | mghdb/mgh239 | 2.5 |
| RI | mghdb/mgh022 | 0.15 |

ECG: electrocardiography; PAP: pulmonary arterial pressure; ART: arterial pressure; CVP: central venous pressure; RI: respiratory impedance.

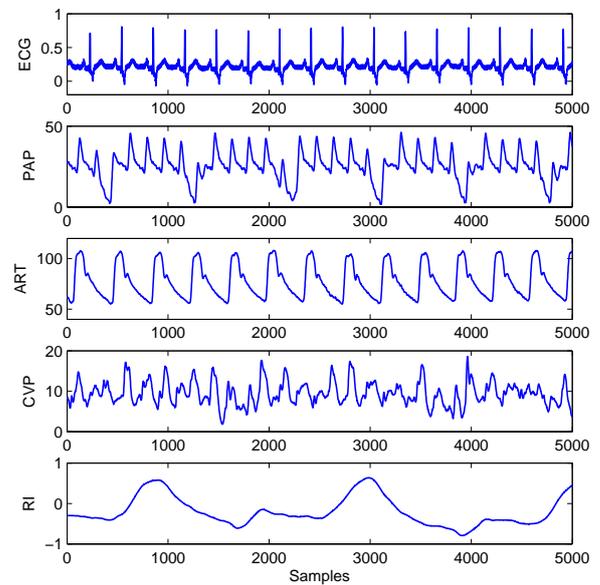

Figure 5. Real-world data samples.

Two main metrics are considered: *energy ratio E* and *data error $\tilde{e}$*, as defined in Section III. In addition, we record the *root mean squared error* (*RMSE*), which is calculated as:

$$RMSE = \sqrt{\frac{\sum_{i=1}^{5000} \tilde{e}_i^{\,2}}{5000}}.$$

Intuitively, smaller energy ratios imply higher energy-efficiency. For comparative study, we evaluate and compare the following algorithms: 1) PAST, 2) AVERAGE, 3) LINEAR, and 4) the proposed prediction algorithm with $K_P$ = 0.6, $K_I$ = 0.4, and $K_D$ = 0.3, which is called PID hereafter for notational convenience. Note that each of the schemes built on these algorithms consists in one particular instance of the proposed approach. We set $M$ = 3 for all simulations.

### B. Results

Fig. 6 shows the energy ratios of each examined scheme on the five datasets. It can be seen that for all datasets, the four schemes can achieve significant energy savings, respectively. For instance, for the RI dataset, both PAST and PID obtain a energy reduction by more than 99%. For all the five datasets, the energy ratios of these two schemes do not exceed 7%, implying energy reductions of more than 93%. The LINEAR scheme performs worst in terms of energy conservation. Regardless of this fact, the resulting energy ratios under LINEAR are all below 41%.

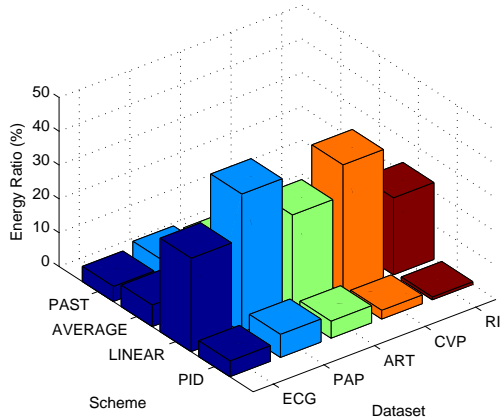

Figure 6. Energy ratios of various schemes for the datasets.

Fig. 7 depicts the data series that will be recorded at the base station (i.e. $\{y_k\}$) under each scheme, along with the corresponding actual samples. A preliminary conclusion from this figure is that the sampled data are followed quite well by all the four schemes, for every dataset. To further examine this issue, we report some subsets of the data errors associated with the four schemes for different datasets in Fig. 8. It is clear that all of the four schemes are able to provide the required information quality. This can be seen from the fact that all the recorded data errors fall within the corresponding error bounds given by $\varepsilon$ in Table II, i.e., the condition (4) is met. The root mean squared errors in all the examined cases are listed in Table III. The results further confirm that the proposed approach is capable of satisfying the user-specified quality requirement over many groups of coefficients. This favorable property makes it quite easy to choose the coefficients for the proposed prediction algorithm.

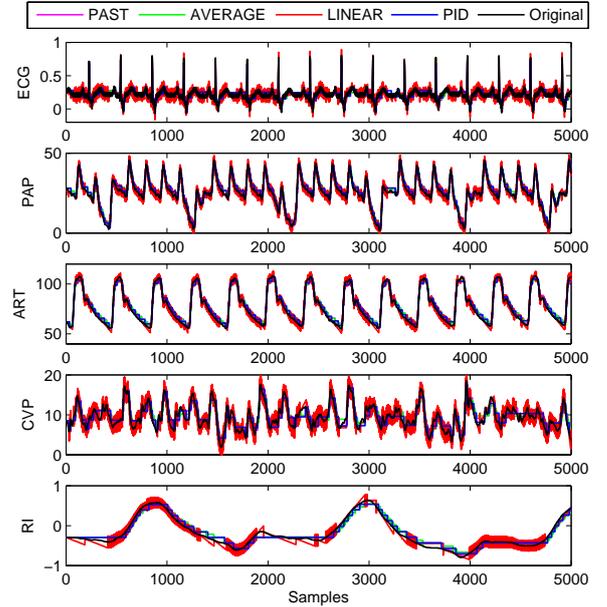

Figure 7. Recorded values and relevant original samples.

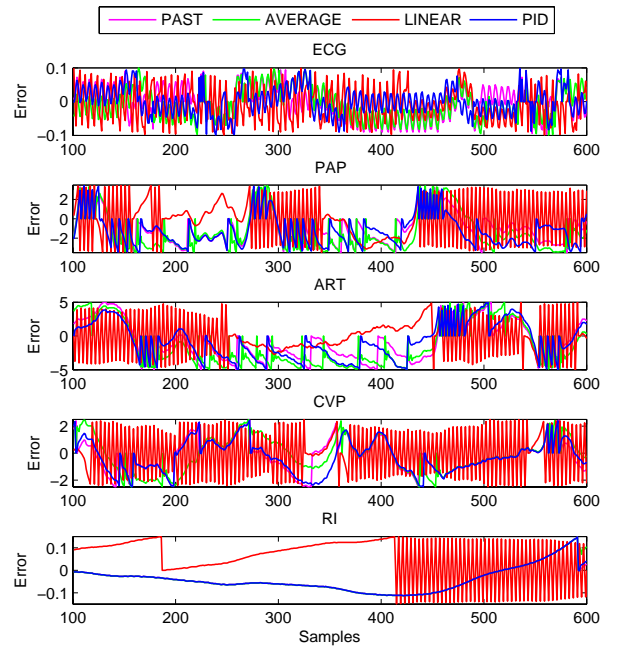

Figure 8. Closeup of data errors.

TABLE III. ROOT MEAN SQUARED ERROR

|     | PAST  | AVERAGE | LINEAR | PID   |
| --- | ----- | ------- | ------ | ----- |
| ECG | 0.043 | 1.845   | 2.735  | 1.225 |
| PAP | 0.046 | 2.338   | 3.446  | 1.431 |
| ART | 0.050 | 1.918   | 2.581  | 1.324 |
| CVP | 0.043 | 1.870   | 2.811  | 1.261 |
| RI  | 0.043 | 1.845   | 2.735  | 1.225 |

To summarize the above results, the proposed approach can be applied to a variety of applications based on wireless body sensors. In addition, our approach is relatively insensitive to changes in its coefficients in that it can deliver quite good performance over a broad range of coefficients, which conforms to the results given in [16]. Depending on the application requirements and the chosen coefficients, our approach can achieve significant energy conservation by 59% to 99% for the datasets examined, while guaranteeing required quality of data all the time.

## VII. CONCLUSIONS

In this paper, we introduced a prediction-based data transmission framework to address the problem of energy conservation in the context of BSNs. The concept of dual prediction has been exploited in the framework. We proposed a generic prediction algorithm for predicting the data sampled by wireless body sensors. This algorithm takes advantage of the well-established PID control technique. Combining the framework and the algorithm yields an effective approach to energy-efficient data transmission. This approach has been designed particularly for BSNs by taking into account their unique characteristics as opposed to conventional WSNs. Simulation results on real-world datasets demonstrate the effectiveness and wide applicability of the approach.

Future work includes extensions to the approach to deal with packet loss and transmission latency, and more extensive simulation and experimental studies.